\documentclass{acm_proc_article-sp}

\begin{document}
\title{GRAPLEr: A Distributed Collaborative Environment for Lake Ecosystem Modeling that Integrates Overlay Networks, High-throughput Computing, and Web Services}
\numberofauthors{5} 
\author{
% 1st. author
\alignauthor
Kensworth Subratie\\
       \affaddr{University of Florida}\\
       \affaddr{Gainesvile, FL, USA}\\
       \email{kcratie@acis.ufl.edu}
% 2nd. author
\alignauthor
Saumitra Aditya\\
       \affaddr{University of Florida}\\
       \affaddr{Gainesvile, FL, USA}\\
       \email{saumitraaditya@acis.ufl.edu}
% 3rd. author
\alignauthor Renato Figueiredo\\
       \affaddr{University of Florida}\\
       \affaddr{Gainesvile, FL, USA}\\
       \email{renato@acis.ufl.edu}
\and  % use '\and' if you need 'another row' of author names
% 4th. author
\alignauthor Cayelan C. Carey\\
       \affaddr{Virginia Tech}\\
       \affaddr{Blacksburg, VA, USA}\\
       \email{cayelan@vt.edu}
% 5th. author
\alignauthor Paul Hanson\\
       \affaddr{University of Wisconsin-Madison}\\
       \affaddr{Madison, WI, USA}\\
       \email{pchanson@wisc.edu}
}
\maketitle

\begin{abstract}
The GLEON Research And PRAGMA Lake Expedition -- GRAPLE -- is a collaborative effort between computer science and lake ecology researchers. It aims to improve our understanding and predictive capacity of the threats to the water quality of our freshwater resources, including climate change. This paper presents GRAPLEr, a distributed computing system used to address the modeling needs of GRAPLE researchers. GRAPLEr integrates and applies overlay virtual network, high-throughput computing, and Web service technologies in a novel way. First, its user-level IP-over-P2P (IPOP) overlay network allows compute and storage resources distributed across independently-administered institutions (including private and public clouds) to be aggregated into a common virtual network, despite the presence of firewalls and network address translators. Second, resources aggregated by the IPOP virtual network run unmodified high-throughput computing middleware (HTCondor) to enable large numbers of model simulations to be executed concurrently across the distributed computing resources. Third, a Web service interface allows end users to submit job requests to the system using client libraries that integrate with the R statistical computing environment. The paper presents the GRAPLEr architecture, describes its implementation and reports on its performance for batches of General Lake Model (GLM) simulations across three cloud infrastructures (University of Florida, CloudLab, and Microsoft Azure). 
\end{abstract}


\section{Introduction}
 
\begin{figure*}
	\centering
		\includegraphics[width=0.8\textwidth]{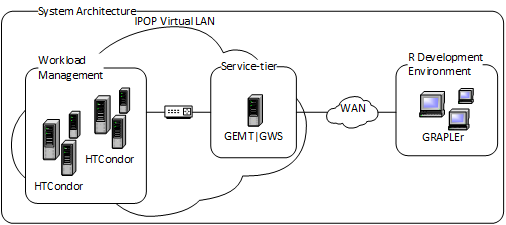}
	\caption{System Architecture (GRAPLEr). Users interact with GRAPLEr using R environments in their desktop (right). The client connects to a Web service tier (GWS) that exposes an endpoint to the public Internet. Job batches are prepared using GEMT and are scheduled to execute in distributed HTCondor resources across an IPOP virtual private network.}
	\label{fig:SystemArchitecture}
\end{figure*}

The GLEON Research And PRAGMA Lake Expedition -- GRAPLE -- aims to improve our understanding and predictive capacity of water quality threats to our freshwater resources, including climate change. It is predicted that climate change will increase water temperatures in many freshwater ecosystems, potentially increasing toxic phytoplankton blooms ~\cite{paerl2008blooms,brookes2011resilience}. Consequently, understanding how altered climate will affect phytoplankton dynamics is paramount for ensuring the long-term sustainability of our freshwater resources. Underlying these consequences are complex physical-biological interactions, such as phytoplankton community structure and biomass responses to short-term weather patterns, multi-year climate cycles, and long-term climate trends~\cite{flynn2005castles}. New data from high-frequency sensor networks (e.g., GLEON) provide easily measured indicators of phytoplankton communities, such as in-situ pigment fluorescence, and show promise for improving predictions of ecosystem-scale wax and wane of phytoplankton blooms~\cite{weathers2013global}. However, translating sensor data to an improved understanding of coupled climate-water quality dynamics requires additional data sources, model development, and synthesis, and it is this type of complex challenge that requires increasing computational capacity for lake modeling.

Searching through the complex response surface associated with multiple environmental starting conditions and phytoplankton traits (model parameters) requires executing and interpreting thousands of simulations, and thus substantial compute resources. Furthermore, the configuration, setup, management and execution of such large batches of simulations is time-consuming, both in terms of computing and human resources.

This puts the computational requirements well beyond the capabilities of any single desktop computer system, and to meet the demands imposed by these simulations it becomes necessary to tap into distributed computing resources. However, distributed computing resources and technologies are typically outside the realm of most freshwater science projects. Designing, assembling, and programming these systems is not trivial, and requires the level of skill typically available to experienced system and software engineers. Consequently, this imposes a barrier to scientists outside information technology and computer science disciplines, and presents challenges to the acceptance of distributed computing as a solution to most lake ecosystem modelers.

GRAPLE is a collaboration between lake ecologists and computer scientists that aims to address this challenge. Through this inter-disciplinary collaboration, we have designed and implemented a distributed system platform that supports compute-intensive model simulations, aggregates resources seamlessly across an overlay network spanning collaborating institutions, and presents intuitive Web service-based interfaces that integrate with existing environments that lake ecologists are used to, such as R.

This paper describes GRAPLEr, a cyberinfrastructure that is unique in how it seamlessly integrates a collection of distributed hardware resources through the IP-over-P2P \cite{ipop06,tincan14} overlay virtual network, supports existing models and the HTCondor distributed computing middleware \cite{thain2005distributed}, and exposes a user-friendly interface that integrates with R-based desktop environments through a Web service. As a multi-tiered distributed solution, GRAPLEr incorporates several components into an application-specific solution. Some of these components are pre-existing solutions which are deployed and configured for our specific uses, while others are specifically developed to address unique needs.

The rest of this paper is organized as follows: Section 2 describes driving science use cases and motivates the need for the GRAPLEr cyberinfrastructure. Section 3 describes the architecture, design, and implementation of GRAPLEr. Section 4 describes a deployment of GRAPLEr and summarizes results from an experiment that evaluates its capabilities and performance. Section 5 discusses related work, and Section 6 concludes the paper.

\section{Architecture and Design}

\subsection{System Architecture (GRAPLEr)}
The system architecture of GRAPLEr is illustrated in Figure~\ref{fig:SystemArchitecture}. Starting from the user-facing components of GRAPLEr, users interact with the system through client-side libraries that are called from an R development environment (e.g., R Studio) running on their personal computer. User requests are mapped by the R library to Application Programming Interface (API) calls that are then sent to the GRAPLEr Web Service (GWS) tier. The GWS tier is responsible for interpreting the user requests, invoking the GRAPLEr Experiment Management Tools (GEMT) to set up a directory structure for model inputs and outputs, and preparing and queuing jobs for submission to the HTCondor pool. The workload management tier is responsible for scheduling and dispatching model simulations across the compute resources, which are interconnected through the IPOP virtual network overlay. These are elaborated below.

\subsection{Overlay Virtual Network (IPOP)}

Rather than investing significant effort in development, porting, and testing new applications and distributed computing middleware, GRAPLEr has focused on an approach in which computing environments are virtualized and can be deployed on-demand on cloud resources. While Virtual Machines (VMs) available in cloud infrastructures provide a basis to address the need for a user-provided software environment, another challenge remains: how to inter-connect VMs deployed across multiple institutions (including private and commercial cloud providers) such that HTCondor and the simulation models work seamlessly? The approach to address this problem is to apply virtualization at the network layer.

The IPOP~\cite{ipop06} overlay virtual network allows GRAPLEr to define and deploy its own virtual private network (VPN) that can span physical and virtual machines distributed across multiple collaborating institutions and commercial clouds. To accomplish this, IPOP captures and injects network traffic via a virtual network interface or ``tap'' device. The ``tap'' device is configured within an isolated virtual private address subnet space. IPOP then encrypts and tunnels virtual network packets through the public Internet. The ``TinCan''~\cite{tincan14} tunnels used by IPOP to carry network traffic use facilities from WebRTC (Web Real-Time Computing) to create end-to-end links that carry virtual IP traffic instead of audio or video. 

To discover and notify peers that are connected to the GRAPLEr ``group VPN'', IPOP uses the eXtensible Messaging and Presence Protocol (XMPP). XMPP messages carry information used to create private tunnels (the fingerprint of an endpoint's public key), as well as network endpoint information (IP address:port pairs that the device is reachable). For nodes behind network address translators (NATs), public-facing address:port endpoints can be discovered using the STUN (Session Traversal Utilities for NAT) protocol, and devices behind symmetric NATs can use TURN (Traversal Using Relays around NAT) to communicate through a relay in the public Internet. Put together, these techniques handle firewalls and NATs transparently to users and applications, and allow for simple configuration of VPN groups via an XMPP server.

\subsection{Workload Management (HTCondor)}
\begin{figure}
	\centering
		\includegraphics[width=0.5\textwidth]{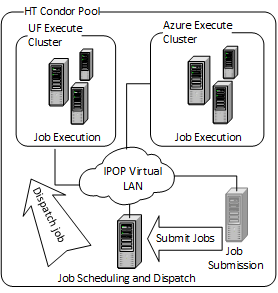}
	\caption{Workload Management (HTCondor). GRAPLEr supports unmodified HTCondor software and configuration to work across multiple sites (e.g., a private cloud at UF and a commercial cloud at MS Azure).}
	\label{fig:CondorPool}
\end{figure}

A key motivation for the use of virtualization technologies, including IPOP, is the ability to integrate existing, unmodified distributed computing middleware. In particular, GRAPLEr integrates HTCondor~\cite{thain2005distributed}, a specialized workload management system for compute-intensive jobs. Like other full-featured batch systems, HTCondor provides a job queueing mechanism, scheduling policy, priority scheme, resource monitoring, and resource management. Users submit their serial or parallel jobs to HTCondor, HTCondor places them into a queue, chooses when and where to run the jobs based upon a policy, carefully monitors their progress, and ultimately informs the user upon completion. Figure~\ref{fig:CondorPool} illustrates the structure of the HTCondor pool that is deployed for GRAPLE.

\subsection{Experiment Management Tools (GEMT)}

\begin{figure}
	\centering
		\includegraphics[width=0.5\textwidth]{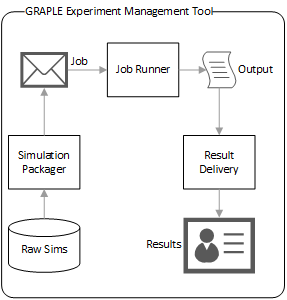}
	\caption{GRAPLEr Experiment Management Tools (GEMT). The GEMT Simulation Packager module takes a specification of the raw simulation inputs and groups them together into jobs; these are dispatched for execution through HTCondor, and their execution at the worker nodes is managed by the GEMT Job Runner module. The Result Delivery GEMT module collates results and presents to the user.}
	\label{fig:GEMT}
\end{figure}

An HTCondor \cite{thain2005distributed} resource pool running across distributed resources connected by IPOP provides a general-purpose capability where it is possible to run a variety of applications from different domains. Furthermore, application-tailored middleware can be layered upon this general-purpose environment to enhance the performance and/or streamline the configuration of simulations on behalf of users. GEMT (Figure~\ref{fig:GEMT}) provides a suite of scripts for designing and automating the tasks associated with running General Lake Model (GLM) based experiments on a very large scale. Here, we use the term \textquotedblleft Experiment\textquotedblright{} to refer to a collection of simulations that address a science use case question, such as determining the effects of climate change on water quality metrics. GEMT provides both the guidelines for the design and layout of individual simulations in the experiment. The primary responsibility of GEMT is to identify and target the task-level parallelism inherent in the experiment by generating proper packaging of executables, inputs, and outputs; furthermore, GEMT seeks to effectively exploit the distributed compute resources across the HTCondor pool by performing operations such as aggregation of multiple simulations into a single HTCondor job, compression of input and output files, and the extraction of selected features from output files. 

For the simulations in an experiment, GEMT defines the naming convention used by the files and directories as well as their layout. The user may interact with GEMT in two possible ways: 1) directly, by using a desktop computer configured with the IPOP overlay software and HTCondor job submission software, or 2) indirectly, by issuing requests against the GRAPLEr Web service. In the former case, once the user has followed the GEMT specification for creating their experiment, executing it and collecting the results becomes a simple matter of invoking two GEMT scripts. However, the user is left the responsibility of deploying and configuring both IPOP and HTCondor locally. Additionally, the user is now a trusted endpoint on the VPN which carries its own security implications. A breach of the user's system is a potential vulnerable point to accessing the VPN. The latter case alleviates the user from both these concerns. This paper focuses on this latter approach, where GEMT scripts are invoked indirectly through the Web service. 

There are three distinct functional modes for GEMT, which pertain to the different phases of the experiment's lifetime. Starting with its invocation, on the submit node, GEMT selects a configurable number of simulations to be grouped as a single HTCondor job. The reason why multiple simulations may be grouped into a single HTCondor job is that, for short-running simulations, the costs of job scheduling and transfer of executables can be significant. By grouping simulations into a single HTCondor job, redundant copies of the input can be eliminated to reduce the bandwidth transfer cost and only a single scheduling decision is needed to dispatch all the simulations in the job. The inputs and executables pertaining to a group of simulations are then compressed and submitted as a job to the HTCondor scheduler for execution. When this job becomes scheduled, GEMT is invoked in its second phase, this time on the HTCondor execute node. The execute-side GEMT script coordinates running each simulation within the job, and preparing the output so it can be returned to the originator. Finally, in its third phase, back on the submit node side, GEMT collates the results of all the jobs that were successful and presents them in a standard format to the end user.

GEMT implements user configurable optimizations to fine tune its operations for individual preferences. It can limit how many simulations are placed in a job, and it will compress these files for transfer. GEMT can also overlap the client side job creation with server side execution to minimize the wait time before results start being produced. These features can be set via a configuration file and combine to provide a simplified mechanism to execute large numbers of simulations.

\subsection{GRAPLEr Web Service (GWS)}

\begin{figure}
       \centering
              \includegraphics[width=0.5\textwidth]{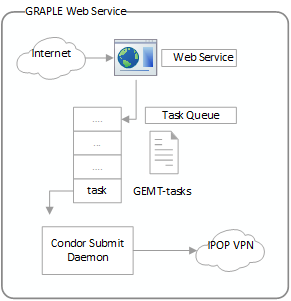}
       \caption{GRAPLEr Web Service (GWS). The GWS is responsible for taking Web service requests from users, interpreting them and creating tasks for remote execution using GEMT.}
       \label{fig:GWS}
\end{figure}

The GWS module, as illustrated in Figure~\ref{fig:GWS}, is a publicly-addressable Web service available on the Internet, and serves as a gateway for users to submit requests to run experiments. GWS acts as a middleware service-tier which exposes an interface to R clients. Requests to run an experiment are made via this interface over the Internet using the HTTP protocol. The functionality provided by GWS is exposed to the R client's user by means of publicly accessible endpoints, each of which is associated with a corresponding method that is invoked in the background. GWS utilizes the functionality of GEMT for simulation processing. GWS generates simulation input files as needed based on the user's request (e.g., to vary air temperature according to a statistical distribution for a climate change scenario), configures and queues jobs, and consolidates and prepares results for download. GWS is co-located in the same host as the GEMT client. This host acts as the submit node to the HTCondor pool, where it monitors job submission and execution. 

 Representational State Transfer or REST, is an architectural style for networked hypermedia applications that is primarily used to build lightweight and scalable Web services. Such a Web service, referred to as RESTful, is stateless with a uniform interface and representation of entities, and communicates with clients via messages and address resources using URIs. GWS implements this paradigm and is designed to treat every job submission independently from any other. Note that there is per-experiment state that is managed by GWS, such as the status of each HTCondor job submitted by the GWS. The state of the experiment is maintained on disk, within the local filesytem, leaving the service itself stateless. GWS implements the public-facing interface using a combination of open-source middleware for Web service processing - Python Flask \cite{Grinberg:2014:FWD:2621997}, and an asynchronous task queue - Python Celery \cite{celery}. The application is hosted using uWSGi (an application deployment solution) and supplemented by a Nginx reverse proxy server to offload the task of serving static files from the application. The employed technology stack facilitates rapid service development and robust deployment.

The GWS workflow begins when a request is received from an R client through the service interface, which is handled by Flask. The request to evaluate a series of simulations can be provided in one of several ways, as discussed in detail in the section covering the R Language Package. However, only data files are accepted as input - no user provided executable binaries or scripts are executed as part of the experiment. A single client-side request can potentially unfold into large numbers (e.g., thousands) of jobs, and GWS places these requests into a Celery task queue for asynchronous processing. Provisioning a task queue allows GWS to decouple the time-consuming processing of the input and task submission to HTCondor from the response to HTTP request.

A 40-character unique identifier (UID) is randomly-generated for each simulation request received by GWS; it is used as an identifier to reference the state of an experiment, and is thus used for any further interactions with the service for a given experiment. Using the UID returned by GRAPLEr, an R client can not only configure the job, but also monitor its status, download outputs, and abort the job. Once the input file has been uploaded to the service, GWS puts the task into the task queue and responds promptly with the UID. Therefore, the latency that the R developer experiences, from the moment the job is submitted to when the UID is received, is minimized. A GWS worker thread then dequeues GEMT tasks from the task queue, and processes the request according to the parameters defined by the user. Figure \ref{fig:GWS} shows the internal architecture and setup of GWS.
A key feature of the GRAPLEr service is to automatically create and configure an experiment by spawning a range of simulation scenarios by varying simulation inputs, based on the user's request and application-specific knowledge. In particular, the service uses application-specific information to identify data in the input file (such as air temperature, or precipitation), and apply transformations to these data (i.e., adding or subtracting an offset to the base value provided by the user) to generate multiple simulation scenarios. GWS removes the burden from the user to generate, schedule, and collate the outputs of thousands of simulations within their own desktops, and allows them to quickly generate experiment scenarios from a high-level description that simply enumerates which input variables to consider, what function to apply to vary them, and how many simulations to create. The user also has the flexibility to retrieve and download only a selected subset of the results back to their desktops, thereby minimizing local storage requirements and data transfer times.

\begin{figure}
	\centering
		\includegraphics[width=0.50\textwidth]{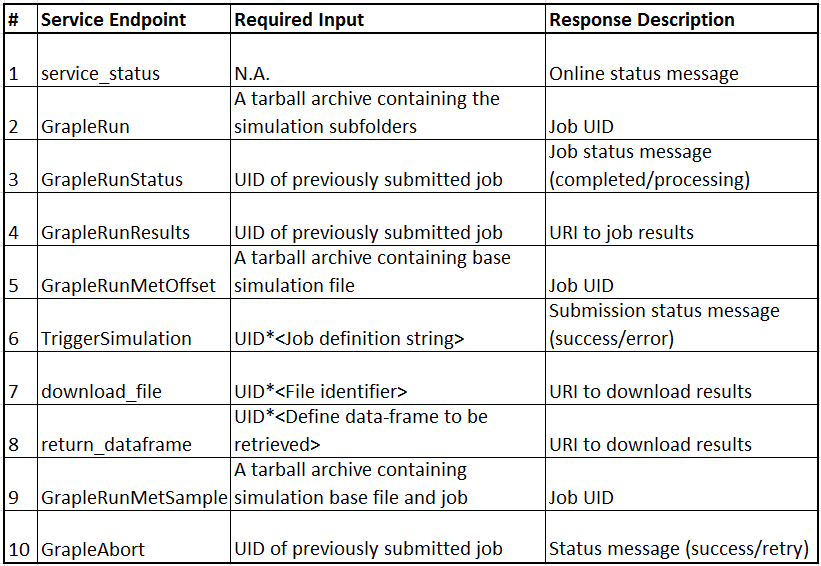}
	\caption{GWS Application Programming Interface (API) Endpoints}
	\label{fig:GWS-endpoints-table}
\end{figure}

To illustrate this feature, consider API endpoint 9 in Figure~\ref{fig:GWS-endpoints-table}. This endpoint exposes a method that enables the user to generate `N' runs from a single baseline set of input files by drawing offsets to input values (e.g., air temperature) from a random distribution. With this API endpoint, the GRAPLEr client can upload a single baseline set of input files, along with a short experiment description file. This file specifies which distribution (random, uniform, binomial, or Poisson) to choose samples from, the number of samples, the variable(s) to be modified, and the operation applied against a variable to each randomly-generated value (add, subtract, multiply, or divide). From this single input and description, GWS generates `N' simulation input files, and calls GEMT Simulation Packager scripts to submit jobs to the HTCondor pool.

\subsection{GRAPLEr R Language Package}

The user-facing component of GRAPLEr is an R package that serves as a thin layer of software between the Web service and the R client development environment (IDE). It exposes an R language application programming interface which can be programmatically consumed by client programs wanting to utilize the GRAPLEr functionality. GRAPLEr is available on github and is installed on the client desktop, where it integrates into the R development environment. It acts as a proxy to translate user commands written in R into Web service calls. It also marshals data between the client and Web service as necessary. The following example illustrates a sequence of three R calls to submit an experiment to a GRAPLEr service running on endpoint graplerURL, from a set of input files placed in sub-directories of a root directory folder on the client-side (expDir), check its status, and download results:

\begin{verbatim}
UID<-GrapleRunExperiment(graplerURL, expDir) 
GrapleCheckExperimentCompletion(graplerURL, UID) 
GrapleGetExperimentResults(graplerURL, UID)
\end{verbatim}

The second example shows how a user can specify a parameter-sweeping simulation with 10,000 simulations which are derived from a baseline set of input files (stored in the simDir directory at the client) by modifying the AirTemp column time series in the GLM meteorological driver input data file met\_hourly.cvs, in the range -10 to 30.

\begin{verbatim}
simDir=C:/Workspace/SimRoot/Sim0
driverFileName=met_hourly.csv
parameterName=AirTemp
startValue=-10 
endValue=30
numberOfIncrements=10000 
expUID<-GrapleRunExperimentSweep(graplerURL, 
 simDir, driverFileName, parameterName, 
 startValue, endValue, numberOfIncrements) 
GrapleCheckExperimentCompletion(graplerURL, expUID) 
GrapleGetExperimentResults(graplerURL, expUID)
\end{verbatim}

To prevent the use of the Web service interface to execute arbitrary code, custom code -- whether binary executables or R scripts -- cannot be sent as part of the simulation requests; instead, users only provide input files and parameters for the GLM simulations. The scenarios that can be run are currently restricted to using GLM tools and our own scripts.

\section{Evaluation}
In this section, we present a quantitative evaluation of a proof-of-concept deployment of GRAPLEr. The goal of this evaluation is to demonstrate the functionality and capabilities of the framework by deploying a large number of simulations to an HTCondor pool. The HTCondor pool is distributed across multiple clouds and connected by the IPOP virtual network overlay. Rather than focusing solely on the reduction in execution times, we evaluate a setup that is representative of an actual deployment composed of execute nodes with varying capabilities.

A GLM simulation is specified by a set of input files, which describe model parameters and time-series data that drive inputs to the simulation, such as air temperature over time, derived from sensor data. The resulting output at the completion of a model run is a netCDF file containing time series of the simulated lake, with many lake variables, such as water temperatures at different lake depths. In our experiments, we use the 1-D GLM Aquatic Eco-Dynamics (AED) model. For a single example GLM-AED simulation of a moderately deep lake run for eight months at an hourly time step, the input folder size was approximately 3 MB, whereas the size of the resulting netCDF file after successful completion of the simulation was 90MB. The test experiment was designed to run reasonably quickly. However, we note that simulations run over decades and with output recorded more frequently may increase simulation time by 1 to 2 orders of magnitude.

We conducted simulation runs on different systems to obtain a range of simulation runtimes. With the baseline parameters, GLM-AED simulation times ranged from the best case of 6 seconds (on a CloudLab system with Intel Xeon CPU E5-2450 with 2.10GHz clock rate and 20MB cache) to 57 seconds (on a University of Florida system with virtualized Intel Xeon CPU X565 with 2.60GHz clock rate and 12MB cache). Note that individual 1-D GLM-AED simulations can be short-running; the GEMT feature of grouping multiple individual simulations into a single HTCondor job leads to increased efficiency.

Description of Experiment setup: The GRAPLEr system deployed for this evaluation was distributed across three sites: University of Florida, NSF CloudLab, and Microsoft Azure. The GWS/GEMT service front-end, HTCondor submit node, and HTC-Central Manager were hosted on virtual machines running in Microsoft's Azure cloud. We deployed three HTC-Execute nodes in total, with 16 cores each. Two nodes were hosted in virtual machines on a VMware ESX server at the University of Florida and one on a physical machine in the CloudLab Apt cluster at University of Utah. All the nodes in this experiment ran Ubuntu-14.04 and HTCondor version 8.2.8; nodes were connected by an IPOP GroupVPN virtual network, version 15.01. Each of the nodes was configured to have 16GB of memory allocated to them.

\begin{figure}
	\centering
		\includegraphics[width=0.50\textwidth]{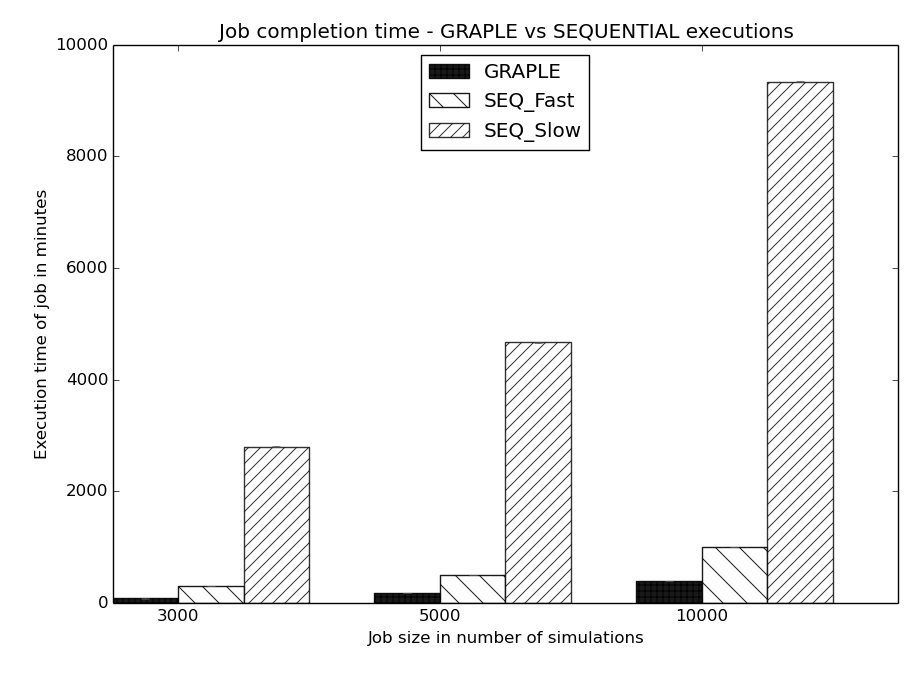}
	\caption{Job runtimes for GRAPLEr HTCondor pool, compared to sequential execution times on CloudLab and UF slots.}
	\label{fig:runtime}
\end{figure}

To conduct the evaluation, we carried out executions of three different experiments containing 3000, 5000 and 10000 simulations of an example lake with varying meteorological input data. Figure \ref{fig:runtime} summarizes the results from this evaluation. As a reference, we also present the estimated best-case sequential execution time on a single, local machine, taken the CloudLab and UF machines as a reference. For 10,000 simulations we achieved a speedup of 2.5 (with respect to sequential execution time of the fast workstation) and 23 (with respect to the sequential execution time at a UF virtual machine).

It is observed that the time taken to complete the job depended greatly on the way simulation tasks were allocated by the HTCondor scheduler. Note that the speedups are relatively modest compared to the best-case baseline, while significant compared to the worst-case baseline. The actual user-perceived speedup would be a function of which desktop environment a user would access the service from. Furthermore, because HTCondor is best-suited for simulations that are individually long-running, the raw user-perceived speedups of GRAPLEr over local execution tend to increase as longer-running simulations are submitted through the service. We expect that, as demand for modeling tools by the lake ecology community increases, so will the complexity, resolution and simulated epochs of climate change scenarios, further motivating users to move from a local processing workflow to remote execution through GRAPLEr.
 
\begin{figure}
	\centering
		\includegraphics[width=0.50\textwidth]{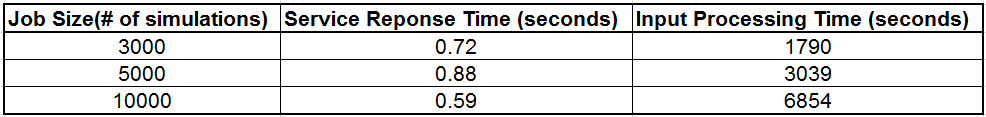}
	\caption{Input handling}
	\label{fig:Results-table}
\end{figure}

Submission of a job to the HTCondor pool involves processing of input (for sweep requests) and packaging of generated simulations into GEMT. In order to evaluate this step we carried out experiments to account for the time taken by GRAPLEr to respond to a request to generate a given number of simulations and submit them for execution. The results are presented in Table \ref{fig:Results-table}. The metric service\_response captures the time taken by GRAPLEr to respond to a request with a UID, which is slightly more than the time required to upload the base\_input . The metric input\_processing captures the time taken to generate and compress all `N' inputs for job submission.

Though not fully explored yet in the design of GRAPLEr, another benefit of remote execution through a Web service interface is the leveraging of storage and data sharing capabilities of the collaborative infrastructure aggregated by distributed resources connected through the IPOP virtual network. For instance, the raw output size of the 10,000 simulation scenario described above is 900 GBytes. By keeping this data on the GRAPLEr cloud and allowing users to share simulation outputs and download selected subsets of the raw data, the service can provide a powerful capability to its end users in enabling large-scale, exploratory scenarios, by both reducing computational time and relaxing local storage requirements at the client side.

\section{Related Work}
Several HTCondor-based high-throughput computing systems have been deployed in support of scientific applications. One representative example is the Open Science Grid (OSG~\cite{osg}), which features a distributed set of HTCondor clusters. In contrast to OSG, which expects each site to run and manage its own HTCondor pool, GRAPLEr allows sites to join a collaborative, distributed cluster by joining its virtual HTCondor pool via the IPOP virtual network overlay. This reduces the barrier to entry for participants to contribute nodes to the network -- e.g., by simply deploying one or more VMs on a private or public cloud. Furthermore, GRAPLEr exposes a domain-tailored Web service interface that lowers the barrier to entry for end users.

The NEWT~\cite{newt} project also provides a RESTful-based Web service interface to High-Performance Computing (HPC) systems. NEWT is focused on providing access to a particular set of resources (NERSC), and does not address the need for a distributed set of (virtualized) computing resources to be interconnected by overlay virtual networks.

\section{Conclusion}
GRAPLEr, a distributed computing system which integrates and applies overlay virtual network, high-throughput computing, and Web service technologies is a novel way to address the modeling needs of interdisciplinary GRAPLE researchers. The system's contribution is its combination of power, flexibility, and simplicity for users who are not software engineering experts but who need to leverage extensive computational resources for scientific research. We have illustrated the system's ability to identify and exploit parallelism inherent in GRAPLE experiments. Additionally, the system scales out, by simply adding additional worker nodes to the pool, to manage both increasingly complex experiments as well as larger number of concurrent users. GRAPLEr is best suited for large numbers of long-running simulations as the distribution and scheduling overhead will increase the running time for such experiments. As lake models demand increased resolution and longer time scales to address climate change scenarios, GRAPLEr provides a platform for the next generation of modeling tools and simulations to better assess and predict the impact to our planet's water systems.

\section{Acknowledgments}
This material is based upon work supported in part by the National Science Foundation under Grants No. 1339737 and 1234983. Any opinions, findings, and conclusions or recommendations expressed in this material are those of the author(s) and do not necessarily reflect the views of the National Science Foundation.

\nocite{*}
\bibliographystyle{abbrv}
\bibliography{pragma-graple} 

\balancecolumns

\end{document}